\documentclass[11pt,twocolumn]{emulateapj_Astroph}

\tightenlines
\usepackage{epstopdf}
\usepackage{ulem} % AV
\usepackage{color} % AV
\usepackage{lineno}

\def\lsim{\;\raise0.3ex\hbox{$<$\kern-0.75em\raise-1.1ex\hbox{$\sim$}}\;}
\def\gsim{\;\raise0.3ex\hbox{$>$\kern-0.75em\raise-1.1ex\hbox{$\sim$}}\;}

\definecolor{purple}{RGB}{200,100,255} %{255,100,20}

\newcommand{\Ntp}{N_\mathrm{TP}}
\newcommand{\nuPk}{\nu_\mathrm{th}}
\newcommand{\nuSSA}{\nu_a}
\newcommand{\EffGeVp}{\epsilon_\mathrm{H}}
\newcommand{\EffGeVhe}{\epsilon_\mathrm{He}}
\newcommand{\EffGeVe}{\epsilon_\mathrm{el}}
\newcommand{\LDi}{L^i_D}
\newcommand{\Dop}{\mathcal{D}}
\newcommand{\CMB}{cosmic microwave background}
\newcommand{\gamray}{$\gamma$-ray}
\newcommand{\Dist}{D_\mathrm{obs}}
\newcommand{\Aobs}{\theta_\mathrm{obs}}
\newcommand{\dMpc}{d_\mathrm{Mpc}}
\newcommand{\LPF}{local plasma frame}
\newcommand{\GRB}{$\gamma$-ray burst}

\newcommand{\UM}{unmodified}

\newcommand{\GBeta}{\gamma(x)\beta(x)}
\newcommand{\GBetaZ}{\gamma_0\beta_0}
\newcommand{\betaX}{\beta(x)}
\newcommand{\gamX}{\gamma(x)}

\newcommand{\LfebDS}{L_\mathrm{DwS}}
\newcommand{\LfebUpS}{L_\mathrm{UpS}}

\newcommand{\Ng}{N_g}
\newcommand{\delMax}{\delta\theta_\mathrm{max}}

\newcommand{\PA}{pitch-angle}

\newcommand{\sigB}{\sigma_B}

\newcommand{\gamZ}{\gamma_0}
\newcommand{\gamT}{\gamma_2}
\newcommand{\betaZ}{\beta_0}
\newcommand{\betaT}{\beta_2}

\newcommand{\Rel}{Relativistic}

\newcommand{\transrel}{trans-rel\-a\-tiv\-is\-tic}
\newcommand{\Transrel}{Trans-rel\-a\-tiv\-is\-tic}
\newcommand{\ultrarel}{ul\-tra-rel\-a\-tiv\-is\-tic}

\newcommand{\Lor}{Lorentz}
\newcommand{\LorT}{Lorentz transformation}

\newcommand{\AZ}{$A/Z$}
\newcommand{\Lmfp}{\lambda_\mathrm{mfp}}
\newcommand{\Fmfp}{f_\mathrm{mfp}}
\newcommand{\SSA}{synchrotron-self-absorption}

\newcommand{\syn}{synchrotron}
\newcommand{\synch}{synchrotron}
\newcommand{\pion}{pion-decay}
\newcommand{\IC}{inverse Compton}
\newcommand{\Hone}{H$^{+}$}
\newcommand{\nHe}{n_\mathrm{He}}

\newcommand{\HeTp}{He$^{2+}\!/p$}
\newcommand{\HeT}{He$^{2+}$}

\newcommand{\HeP}{\mathrm{He}^{+2}\!/p}
\newcommand{\aaT}{{\bf A}}
\newcommand{\bBB}{{\bf B}}
\newcommand{\cCC}{{\bf C}}
\newcommand{\dDD}{{\bf D}}
\newcommand{\eEE}{{\bf E}}
\newcommand{\fFF}{{\bf F}}
\newcommand{\gGG}{{\bf G}}
\newcommand{\hHH}{{\bf H}}
\newcommand{\iII}{{\bf I}}
\newcommand{\jJJ}{{\bf J}}
\newcommand{\degg}{^\circ}

\newcommand{\xx}[1]{\!\times\!10^{#1}}

\newcommand{\pcc}{cm$^{-3}$}

\newcommand{\muG}{$\mu$G}

\newcommand{\RH}{Rankine-Hugoniot}
\newcommand{\rRH}{R_\mathrm{RH}}

\newcommand{\mfp}{mean free path}

\newcommand{\pic}{particle-in-cell}

\newcommand{\rg}{r_g}
\newcommand{\rgz}{r_{g0}}

\newcommand{\gyrotime}{\tau_g}
\newcommand{\deltime}{\delta t}
\newcommand{\etamfp}{\eta_\mathrm{mfp}}
\newcommand{\EnTran}{f_\mathrm{ion}}

\newcommand{\Rtot}{R_\mathrm{tot}}
\newcommand{\Rsub}{R_\mathrm{sub}}

\newcommand{\Fen}{F_\mathrm{en}}
\newcommand{\FenZ}{F_\mathrm{en}^0}

\newcommand{\nonrel}{non-relativistic}

\newcommand{\rel}{relativistic}
\newcommand{\Emax}{E_\mathrm{max}}
\newcommand{\pmax}{p_\mathrm{max}}

\newcommand{\mc}{Monte Carlo}
\newcommand{\MC}{Monte Carlo}
\newcommand{\DSA}{diffusive shock acceleration}
\newcommand{\Facc}{Fermi acceleration}
\newcommand{\FoFSA}{first-order Fermi shock acceleration}

\newcommand{\TP}{test-particle}
\newcommand{\NL}{nonlinear}
\newcommand{\SCly}{self-consistently}
\newcommand{\SC}{self-consistent}

\renewcommand{\baselinestretch}{1}
\def\aap{Astronomy and Astrophysics}

\def\apjl{The Astrophysical Journal Letters}

\newcount\listnorom
\newcommand\listromanDE{\global\advance \listnorom by 1
{\lowercase\expandafter{(\romannumeral\listnorom)}\ }}
\newcommand\newlistroman{\listnorom=0}

\newcount\listnumber
\newcommand\listDE{\global\advance \listnumber by 1
{\lowercase\expandafter{(\number\listnumber)}\ }}

\newcount\Inum
\newcount\IInum
\newcount\IIInum
\newcount\IVnum
\def\I{\global\multiply\IInum by 0 \global\multiply\IIInum by 0
            \global\multiply\IVnum by 0 \global\advance \Inum by 1
            {\the\Inum. }}
\def\II{\global\multiply\IIInum by 0\global\multiply\IVnum by 0
       \global\advance \IInum by 1 {\the\Inum.\the\IInum. }}
\def\III{\global\multiply\IVnum by 0\global\advance \IIInum by 1
            {\the\Inum.\the\IInum.\the\IIInum. }}
\def\IV{\global\advance \IVnum by 1
            {\the\IVnum. }}
\shorttitle{Particle Acceleration in Relativistic shocks}
\shortauthors{Warren, Ellison, Bykov, Lee}

\begin{document}
%\medskip

\title{Electron and Ion Acceleration in Relativistic shocks with Applications to GRB Afterglows} 

\vskip24pt

\author{Donald C. Warren,\altaffilmark{1}
Donald C. Ellison,\altaffilmark{2}
Andrei M. Bykov,\altaffilmark{3,4,5} and
Shiu-Hang Lee\altaffilmark{6}}

\altaffiltext{1}{Astrophysical Big Bang Laboratory, RIKEN, Saitama 351-0198, Japan; donald.warren@riken.jp}

\altaffiltext{2}{Physics Department, North Carolina State
University, Box 8202, Raleigh, NC 27695, U.S.A.;
don\_ellison@ncsu.edu}

\altaffiltext{3}{Ioffe Institute for Physics and Technology, 194021
St. Petersburg, Russia; ambykov@yahoo.com}

\altaffiltext{4}{International Space Science Institute, Bern, Switzerland}

\altaffiltext{5}{St.Petersburg Polytechnic University, 
St. Petersburg, Russia}

\altaffiltext{6}{Japan Aerospace Exploration Agency (JAXA),
Institute of Space and Astronautical Science, Kanagawa 252-5210, Japan; slee@astro.isas.jaxa.jp}

%PACS numbers: 94.20.Wc, 98.38.Mz, 98.70.Sa

\begin{abstract}
We have modeled the simultaneous \FoFSA\ of protons, electrons, and helium nuclei by \rel\ shocks. 
By parameterizing the particle diffusion, our steady-state Monte Carlo simulation allows us to follow particles from  particle injection at  \nonrel\ thermal energies to above PeV energies, including 
the \NL\ smoothing of the shock structure due to cosmic-ray (CR) backpressure.
We observe the mass-to-charge (\AZ) enhancement effect believed to occur in efficient \Facc\ in \nonrel\ shocks and we parameterize the transfer of ion energy to electrons seen in \pic\ (PIC) simulations. 
For a given set of environmental and model parameters, 
the \mc\ simulation determines the absolute normalization of the particle distributions and the resulting \synch, \IC, and \pion\ emission in a largely \SC\ manner. 
The simulation is flexible and can be readily used with a wide range of parameters typical of \GRB\ (GRB) afterglows. We describe some preliminary results for photon emission from shocks of different \Lor\ factors and outline how the \mc\ simulation can be generalized and coupled to hydrodynamic simulations of GRB blast waves. 
We assume Bohm diffusion for simplicity but emphasize that the \NL\ effects we describe stem mainly from an extended shock precursor where higher energy particles diffuse further upstream. Quantitative differences will occur with different diffusion models, particularly for the maximum CR energy and photon emission, but these \NL\ effects should be qualitatively similar as long as the scattering \mfp\ is an increasing function of momentum. 
\end{abstract}

\section{Introduction}
Efficient \FoFSA\ (also called \DSA) is often suggested as a likely mechanism for converting the bulk kinetic energy of \rel\ plasma flows into individual particle energy 
\citep[e.g.,][]{BykovTreumann2011,BykovEtal2012}. However, many aspects of particle acceleration in \rel\ shocks remain uncertain because of the inherent complexity  of the process. The particle distributions are  highly anisotropic and the 
magnetic turbulence,  essential for acceleration to occur, must be self-generated and is extremely difficult to 
characterize \citep[][]{LP2003,NO2006,LP2010,RevilleBell2014,LPGP2014}. 
These roadblocks can be overcome with \pic\ (PIC) simulations and intensive work has been done in this area 
\citep[e.g.,][]{NishikawaEtal2007,SironiSpit2011}. However, current PIC simulations are computationally costly and have a limited dynamic range.
The \transrel\  regime, which may be important for 
GRB afterglows \citep[e.g.,][]{Meszaros2006,AckermannEtal2013} and some types of 
supernovae \citep[e.g.,][]{Chakraborti2011}, is less well explored either analytically or using PIC simulations 
\citep[see, however,][]{Casse2013}.

In this paper we model the \NL\ acceleration of electrons and 
ions (protons and \HeT) at \rel\ collisionless shocks using a \mc\ simulation of \FoFSA\ \citep*[e.g.,][]{ED2002,EWB2013}. 
The steady-state \mc\ simulation parameterizes magnetic turbulence generation and particle diffusion, important approximations but ones that allow a large dynamic range;  the simultaneous acceleration of ions and electrons (along with the radiation they produce); and a \SC\ determination of the shock structure.
We believe this is the first attempt, apart from PIC simulations, to include electrons \SCly\  with ions in a \NL, \rel, \Facc.  

While it is  well known that \NL\ \Facc\ should preferentially inject and accelerate  high mass-to-charge particles compared to protons in \nonrel\ shocks \citep[e.g.,][]{EJE1981,Eichler84,JE91}, to our knowledge this process has not been investigated in \rel\ or \transrel\ shocks until now. 
The \nonrel\ ``$A/Z$" effect ($A$ is the mass in units of the proton mass $m_p$ and $Z$ is the charge in units of the electron charge $e$)
has been shown to be consistent with observations of diffuse ions accelerated at the Earth bow shock \citep[][]{EMP90} and has been used to model the shock acceleration of interstellar gas and dust, matching important aspects of the galactic cosmic-ray abundances observed at Earth 
\citep[e.g.,][]{EDM97,MDE97,ME1999,RauchErr2010,BinnsICRC2013}. 

\newlistroman

The \AZ\ enhancement we model is a purely kinematic effect. It depends on the following assumptions:
\listromanDE 
the acceleration process is efficient enough so the backpressure from accelerated particles noticeably modifies the shock precursor,
\listromanDE 
all particles have a scattering mean-free-path of the approximate form
\begin{equation} \label{eq:mfp}
\Lmfp = \etamfp \Fmfp 
\ ,
\end{equation}
where $\Fmfp$ is an increasing function of local frame momentum $p$, as is generally assumed, and 
\listromanDE 
the normalization parameter $\etamfp$ setting the ``strength" of scattering is 
similar for all particle species.

The simplest assumption for particle diffusion is that $\Fmfp$ equals the gyroradius, i.e., $\Fmfp = \rg =pc/(ZeB)$. Then 
\begin{equation} \label{eq:mfp_two}
\Lmfp = \etamfp \rg
\ ,
\end{equation}
and the precursor diffusion length for species $i$ is
\begin{equation}
\LDi \propto \etamfp \gamma_i (A/Z) v_i^2
\ .
\end{equation}
Here $\gamma_i$ ($v_i$) is the \Lor\ factor (velocity) for species $i$, $c$ is the speed of light, and $B$ is the background magnetic field in Gauss used to scale $\rg$.
If $v_i \sim c$, then $\gamma \propto p/A$,
\begin{equation}
\LDi \propto \etamfp (A/Z) (p/A)
\ ,
\end{equation}
and the precursor diffusion length for different \AZ\ ions,  at the same momentum per nucleon, scales as \AZ.

Therefore, if $\etamfp$ is similar for all species, at the same $p/A$, high \AZ\ particles will diffuse further into the upstream region, `feel' a larger effective compression ratio in the modified shock structure than low \AZ\ particles, and gain a larger momentum boost in the next shock crossing. 
Since $\LDi$ increases with $p/A$, the modified shock precursor produces a distinctive concave spectral shape \citep[e.g.,][]{EE84} with  high \AZ\ particles having  a harder spectrum at any $p/A$ than low 
\AZ\ particles (see Fig.~\ref{fig:dndp_AZ} below).
It is important to note that equation~(\ref{eq:mfp_two}) is  presented only as a simple baseline scattering mode. Small-scale turbulence, as generated by the Weibel instability at \rel\ 
shocks \citep[e.g.,][]{Nishikawa2006,PPL2011,PPL2013}, in general leads to a different $\Lmfp \propto p^2$ dependence \citep[e.g.,][]{Jokipii1971}. 
In fact,  the scattering process in relativistic shocks is almost certain to be more complicated than any simple power-law dependence for $\Lmfp$, and $\etamfp$ may be expected to have a momentum dependence as well
\citep[e.g.,][]{Achterberg2001,KR2010,SSA2013,LPGP2014}.
Our current results, with $\Lmfp \propto \rg$, display the essential physical effects that come about from the development of an extended shock precursor. As long as $\Lmfp$ is an increasing function of $p$, and \Facc\ is efficient, the \NL\ effects we describe should not depend qualitatively on the diffusion coefficient. Quantitatively, the momentum dependence of $\Lmfp$ can strongly influence the maximum CR energy a given shock can produce and this, in turn, will influence the photon production. We are currently generalizing our \MC\ simulation to allow for a more complicated parameterization of $\Lmfp$ over large energy and length scales. These future results will be compared against PIC 
simulations \citep[e.g.,][]{SSA2013}.
The results presented here  provide a benchmark for future comparisons.

While the \AZ\ effect we describe can enhance the injection and acceleration of heavy ions compared to protons, for electrons with $A/Z \simeq 5.45\xx{-4}$ it acts strongly in the opposite fashion.
If only this kinematic effect is considered with equation~(\ref{eq:mfp_two}), electrons will be dramatically less efficiently injected and accelerated than protons in \NL\ \Facc.
 
Electron injection was considered in a \nonrel\ \mc\ code similar to the one we use here in \citet{BaringEtal99}. In that paper, in order to overcome the dramatic \AZ\ effect and allow electrons to be injected and accelerated with efficiencies large enough to be consistent with \syn\ and IC radiation observed in young supernova remnants (SNRs), the electron  $\Lmfp$ was set equal to a constant below some momentum, i.e., changing $\etamfp$ selectively for electrons.
This modification gave low energy electrons a larger mean free path than equation~(\ref{eq:mfp_two}) would produce and allowed them to diffuse far enough upstream to overcome the shock smoothing effects. It was argued in \citet{BaringEtal99} that this simple modification was reasonably consistent with an electron injection model developed by \citet{Levinson1992}.

Here, we adopt a different approach. We keep equation~(\ref{eq:mfp_two}), but transfer some fraction of the ram kinetic energy from ions to electrons as the particles first cross the viscous subshock.
We note that a ``sharing" of energy between ions and electrons is 
clearly seen in recent PIC  \rel\ shock simulations  
\citep[i.e.,][]{SironiSpit2011,SSA2013}, and \citet*{PPL2013} give   
an analytical treatment of electromagnetic instabilities 
transferring energy to electrons in the precursor of relativistic shocks.

In the PIC simulations, electrons are heated in the precursor
by interacting with turbulence generated mainly by backstreaming protons and obtain near equipartition with the protons  before crossing the subshock. We mimic this effect by transferring 
a set fraction, $\EnTran$, of ion energy to electrons as particles first cross the subshock.
While this simple energy transfer model is clearly an approximation, we feel it affords a straightforward way of using plasma physics information obtained from computationally intensive PIC simulations in a calculation that can model particle acceleration and photon emission consistent with the production of high-energy 
cosmic rays in \rel\ shocks.\footnote{Other descriptions of electron heating in \rel\ plasmas include \citet{Gedalin2008}, who investigate the effects of a cross-shock potential, and \citet{Kumar2015}, who explore the electron heating behavior of Weibel-induced current filaments in a 2D PIC simulation.}

We recognize, of course, that collisionless shock formation, and particle injection and acceleration, are determined by more than kinematics alone. The self generation of magnetic turbulence is critical to the 
process \citep[e.g.,][]{BO98,LP2003}, particularly for \rel\ shocks, and, as just mentioned,  energy can be transferred between 
electrons and protons by wave-particle interactions that are essentially independent of the kinematics. 
Nevertheless, if \Facc\ is efficient, basic momentum and energy conservation demands that kinematics be taken into account and the shock precursor must be modified by the backpressure of accelerated particles. 

In contrast to \nonrel\ shocks, where shock acceleration can be directly tested against spacecraft observations \citep[e.g.,][]{EMP90,BOEF97}, 
\Facc\  in \rel\ shocks is far less certain. \Rel\ shocks cannot be directly observed with spacecraft, as can \nonrel, heliospheric shocks, and the highly anisotropic particle distributions intrinsic to \rel\ shocks make the self-generation of magnetic turbulence far more difficult to describe analytically. 
Despite this difficulty, intensive work continues \citep[e.g.,][]{LP2010,LP2011,PPL2013}.
Furthermore, and again in contrast to \nonrel\ shocks, the predictions for particle spectra and, therefore, photon signatures 
from \rel\ shocks are highly uncertain. Sites harboring \rel\ shocks, such as GRBs, can often be successfully
modeled with alternative acceleration mechanisms 
\citep[e.g., magnetic reconnection 
in the case of GRBs;][]{McKinney2012,SironiPG2015}, 
weakening the link between \Facc\ theory and observation.

Despite this uncertainty, there is compelling evidence, primarily from PIC simulations 
\citep[e.g.,][]{Hoshino92,Kato2007,
SironiSpit2009,KeshetEtal2009,NishikawaEtal2011,SironiSpit2011,Stockem2012}, that \rel\ shocks do accelerate electrons and protons beyond the initial kinematic boost from a single shock crossing.
These simulations also highlight the role of the magnetization parameter, $\sigB$, in \Facc, where
\begin{equation} \label{eq:sigma}
\sigB = \frac{B_0^2}{4 \pi n_0 m_p c^2}
\ ,
\end{equation}
$B_0$ is the upstream field, and $n_0$ is the number density 
\citep[see][ for an  alternative definition of $\sigB$]{BykovTreumann2011}.

\citet{SSA2013} \citep[see also][]{Haugbolle2011}
show that  perpendicular electron--ion  shocks with Lorentz factors $\gamZ \lsim 150$  inject and accelerate electrons and ions efficiently when  $\sigB\, \lsim 3\xx{-5}$. 
In these weakly magnetized plasmas, the self-generated turbulent field dominates the uniform $B$ and the shock obliquity ceases to be important, an assumption often made in shock acceleration studies and one we make here with our plane-parallel shock assumption (c.f., Fig.~\ref{fig:Jet_detail}). Quantifying this lack of dependence on the obliquity is particularly important since \ultrarel\ shocks are essentially always highly oblique---the $B$-field component along the shock face can be highly compressed---and strongly magnetized oblique shocks are less  able to inject and accelerate particles.
We note that for typical interstellar medium (ISM) conditions,
$B_0 \sim 3$\,\muG, $n_0 \sim 0.03$\,\pcc, and so $\sigB \sim 10^{-9}$. This value is orders of magnitude below the threshold reported by 
\citet{SSA2013}, suggesting that \ultrarel\ shocks propagating in the normal ISM may be able to inject and accelerate electrons and ions far more efficiently than previously believed, regardless of the shock obliquity.
We note that for \nonrel\ shocks the plasma $\beta \equiv n_0 k_B T_0/(B_0^2/8\pi)$ is the relevant parameter rather than
equation~(\ref{eq:sigma}) and oblique geometry may  be important for typical ISM parameters \citep[e.g.,][]{Orlando2011}. Here $k_B$ is Boltzmann's constant and $T_0$ is the ambient unshocked temperature.

Given that weakly magnetized \rel\ shocks can inject particles, the maximum energy these particles obtain in a given shock remains uncertain, although arguments presented by \citet{SSA2013} suggest 
that the acceleration reaches a maximum, $\Emax$, where
$
\Emax/(\gamZ m_p c^2) \sim \sigB^{-1/4}.
$
While this can be a substantial energy (for $\sigB = 3\xx{-5}$, $\Emax \sim 0.2$\,TeV), it is well below what is often 
assumed in suggesting that \rel\ shocks may produce 
ultra-high-energy cosmic rays \citep[e.g.,][]{KW2005}. Here we simply parameterize the maximum particle energy by setting a maximum shock size with a free escape boundary (FEB).\footnote{A FEB is a position beyond which particles are assumed to decouple from the shock. Any actual shock will be finite in extent and at some point high-energy particles will obtain diffusion lengths comparable to the shock size and stream freely away. Our FEBs model a finite shock size within our steady-state, 
plane-parallel approximation.}

Next we describe the generalization of the \mc\ code used in \citet{EWB2013} to include the injection and acceleration of electrons simultaneously with ions. The \NL\ shock structure is calculated including thermal leakage injection and the backreaction from all species.
Full particle spectra are determined at various positions relative to the subshock, along with the distributions of particles that escape at upstream and downstream FEBs.
For given values of the ambient density, magnetic field, and background photon field, we calculate the \syn\ emission using 
equation~(6.7a) in \citet{Rybicki79}, 
the \IC\ (IC) emission using equation~(9) in \citet{Jones68}, 
and the \pion\ emission using parameterizations 
given by \citet{Kamae06,Kamae2007} and \citet{Kelner2009}. For the \pion\ from \HeT, we use the scaling relation given in 
\citet{BaringEtal99}. Once the emission is determined in the local frame it is transformed to the observer (i.e., ISM) frame.

We note that radiation losses for electrons are \hbox{only} considered during the acceleration process---once accelerated, the electrons radiate without accounting for further losses. This so-called `thin target' approximation, where the radiation length is assumed to be larger than the region between the upstream and downstream FEBs, is adequate for the steady-state examples given here and can be relaxed in models of evolving GRBs \citep[see][]{Warren2015dis}.

\section{Model} \label{sec:Model}
We assume the basic \mc\ scattering model, 
as described in \citet{EWB2013} for protons, applies equally to electrons and heavy ions.
If so, all that is needed to describe the injection and acceleration of electrons is a parameter describing the transfer of energy from ions to electrons, mimicking the effect  seen in PIC simulations of \rel\ shocks \citep[see][]{SironiSpit2011}.

Full details of the particle scattering model, 
thermal leakage injection, and the method for obtaining a \SC\ shock precursor structure when \Facc\ and particle escape occur are given in Section~2 of 
\citet{EWB2013}.
In \citet{EWB2013} we also fully explain the caveats needed when applying the \mc\ model to \rel\ shocks and show how the approximations required in the \mc\ model compare with previous \mc\ work and with more fundamental PIC simulations.

We describe particle transport by assuming the mean free path, $\Lmfp$, is given by equation~(\ref{eq:mfp_two}) 
with $\etamfp=1$, i.e., Bohm diffusion.\footnote{We note that since all lengths in the 
steady-state code are scaled with $\rgz=\etamfp m_p u_0 c/(e B_0)$, where $u_0$ is the shock speed, our results are independent of $\etamfp$ except for the absolute normalization of the particle spectra, i.e., the number of particles within a physical region scales as $\etamfp$. 
In our plane-parallel approximation, the field throughout the shock retains the far upstream value $B_0$.}
As mentioned above, while there is some theoretical and observational support for Bohm diffusion from X-ray 
afterglows \citep[e.g.,][]{SagiNakar2012}, most  theoretical work suggests a stronger momentum dependence, i.e., 
$\Lmfp \propto p^2$ \citep[e.g.,][]{PPL2011}. We use
equation~~(\ref{eq:mfp_two}) for convenience and as a baseline for future work which will assume more complicated (and, hopefully, more realistic) forms for $\Lmfp$.
The particle is moved for a time $\deltime \ll t_c$, where
$t_c=\Lmfp/v_i$ is the ``collision time," i.e., the average time measured in the  local frame needed for the particle to accumulate deflections on the order of $90\degg$.
The second scattering parameter, $\Ng$, determines the ``fineness" of scattering through an equation for the maximum deflection, $\delMax$, a particle can experience in a \PA\  interaction event after each $\deltime$, i.e.,
\begin{equation}\label{eq:Tmax}
\delMax = \sqrt{12 \pi /(\etamfp \Ng)}
\ .
\end{equation}
Here, $\Ng= \gyrotime /\deltime$ is the number of 
gyro-time segments $\deltime$ dividing a gyro-period $\gyrotime=2\pi\rg/v_i$,
and we note that equation~(\ref{eq:Tmax}) applies even if particles move rapidly between inertial frames, as is normally the case for \rel\ shocks.
In each scattering event, the scattering is assumed to be isotropic and elastic in the local plasma frame.

Large values of $\Ng$ imply fine scattering while small values imply that the particle momentum will suffer relatively large deviations in direction in each \PA\ scattering event 
\citep[e.g.,][]{SummerlinBaring2012}.
In all examples here, $N_g$ is large enough to 
saturate and produce fine-scattering results that do not change substantially as $N_g$ is increased further.\footnote{We note that all of our fully \rel, unmodified shock examples yield the well known power law $dN/dp \propto p^{-2.23}$ \citep[e.g.,][]{BO98,KirkEtal2000,KW2005} 
(see Fig.~\ref{fig:UM_gam10} below). Harder spectra can be obtained with large-angle scattering, i.e., small values of $\Ng$ \citep[e.g.,][]{ED2004,SummerlinBaring2012}, or other ad hoc assumptions \citep[e.g.,][]{Schlickeiser2015} but care must be taken to account for NL effects when the power law index is less than 2 \citep[see figure~13 in][]{EWB2013}.} 
It is important to note that while we can simply parameterize particle transport with the assumption given by equation~(\ref{eq:mfp_two}) and the two parameters $\Lmfp$ and $\Ng$, this assumes that magnetic fluctuations with correlation lengths on the order of $L_c = 2 \pi \etamfp \rg /\Ng$ exist with sufficient power to produce this scattering throughout the shock. We make no claim of \SCly\ determining the magnetic turbulence needed to produce the diffusion implied by equation~(\ref{eq:mfp_two}).\footnote{We note that a  \NL, \nonrel\ \mc\ code which does include the self-generation of magnetic turbulence has been 
developed \citep[e.g.,][]{VBE2008,VBE2009,Bykov3inst2014}.} 

All of the above is applied equally to electrons and ions and this parameterization yields injection rates and acceleration efficiencies that depend on the $A/Z$ ratio.
We note that particle injection from shock-heated thermal 
particles (our thermal leakage injection model) occurs directly from the above assumptions.  Regardless of the shock speed, virtually all cold, thermal, unshocked particles  obtain a velocity greater than the downstream bulk flow speed when they first scatter in the downstream region  and the flow becomes subsonic. Depending on their angular distribution, these particles have a finite probability of scattering back upstream where they will be further accelerated. 
In the \mc\ code, the injection efficiency is determined stochastically as some particles manage to scatter back upstream with an angular distribution determined solely from the scattering model with no additional assumptions or parameters other than that the subshock is assumed to be transparent for back-scattering particles.

To this we add one additional parameter, $\EnTran$, i.e., the fraction of far upstream ion ram kinetic energy transferred to electrons.  The ion ram kinetic energy is defined as
$(\gamZ - 1) m_i c^2$,
where $m_i$ is the ion mass, $\gamZ=[1 -(u_0/c)^2]^{-1/2}$ is the far upstream shock Lorentz factor, and $u_0$ is the far upstream (i.e., unmodified) shock speed.
When an ion crosses the subshock from upstream to downstream for the first time, energy
equal to 
$\EnTran (\gamZ - 1) m_i c^2$
is removed from it. 
The total ion energy transferred is 
$\EnTran \sum_i(\gamZ - 1) N_i m_i c^2$,
where the sum is over the ion species and $N_i$ is the number of 
$i$-ions injected far upstream.
This ion energy
is divided equally among electrons and added to their energy  as they cross the subshock into the downstream region for the first time.
The parameter $\EnTran$ can increase electron injection substantially as we show below.

\begin{figure}
\epsscale{1.0}
\plotone{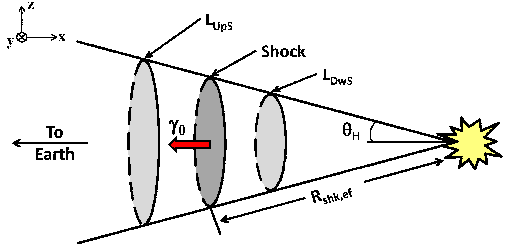}           % Fig 1 
\caption{Schematic, not-to-scale representation of a \rel\ shock embedded in a conical jet, propagating with a Lorentz factor $\gamZ$ into material at rest.  The region of interest is delimited by $L_\mathrm{UpS}$ and $L_\mathrm{DwS}$; outside of this region, we assume the particles decouple from the shock and we ignore any emission resulting from them.
The jet opening half-angle is $\theta_{H}$, and the shock has propagated a distance $R_\mathrm{shk,ef}$ in the explosion (or ISM) frame. We only consider emission directed along the jet to an observer at Earth.
\label{fig:Jet_cone}}
\end{figure}

\begin{figure}
\epsscale{1.0}
\plotone{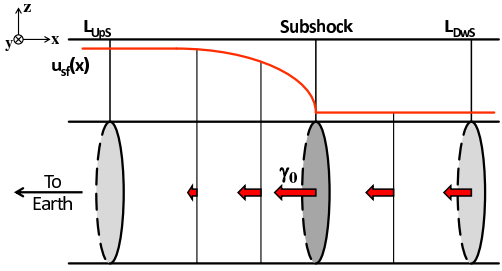}           % Fig 2
\caption{This detail focuses on the volume between $L_\mathrm{UpS}$ and $L_\mathrm{DwS}$.  The upper panel shows the shock-frame plasma velocity profile.  The lower panel shows the ISM-frame velocity at selected points in the shock structure, varying from 0 for $x<\LfebUpS$ to $u_0$ just upstream from the subshock at $x=0$. 
Note that we assume the shock is locally plane and that $L_\mathrm{UpS}$ and $L_\mathrm{DwS}$ are small compared to $R_\mathrm{shk,ef}$ as illustrated in Fig.~\ref{fig:Jet_cone}.
\label{fig:Jet_detail}}
\end{figure}

\begin{figure}
\epsscale{1.00}
\plotone{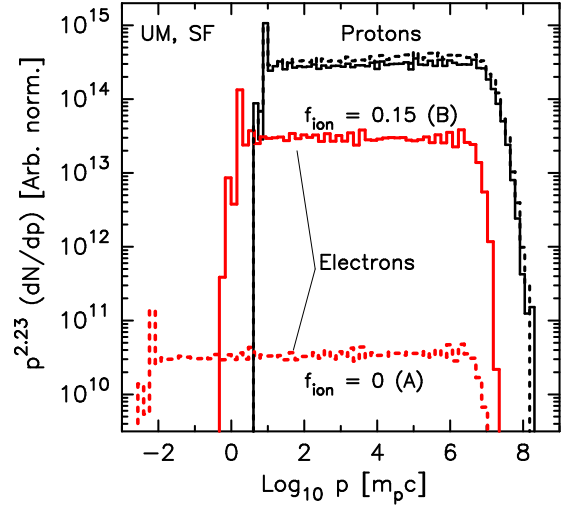}           % Fig 3      
\caption{Protons (black curves) and electrons (red curves) from UM shocks with different $\EnTran$ as indicated. These spectra, multiplied by $p^{2.23}$, are calculated downstream from the shock, in the shock frame, and have arbitrary overall normalization although the relative normalization between electrons and protons is absolute. An upstream FEB of 
$\LfebUpS=-10^4\,\rgz$ was used with no downstream FEB, i.e., a probaility-of-return calculation was used to simulate an infinite downstream region. In Table~\ref{tab:param} the $\EnTran=0$ case is Model \aaT\ and the $\EnTran=0.15$ case is Model \bBB.
\label{fig:UM_gam10}}
\end{figure}

\begin{figure}
\epsscale{1.00}
\plotone{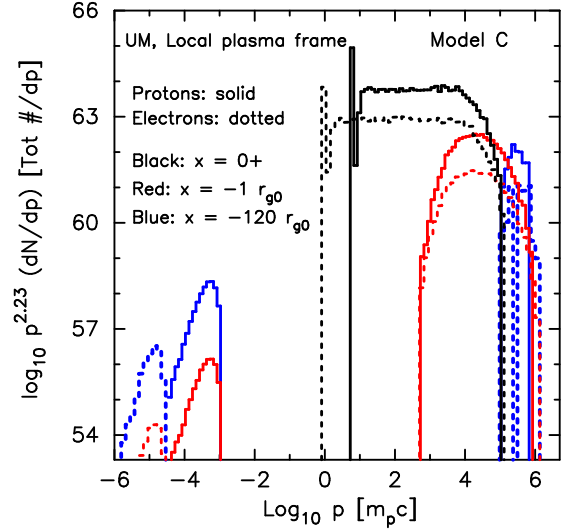}           % Fig 4   
\caption{Proton and electron spectra (as labeled and multiplied by $p^{2.23}$) with $\EnTran=0.15$. These spectra are measured in the local plasma frame and are normalized to the total number of particles in a given region, as indicated in  Fig.~\ref{fig:Jet_detail}. The shock acceleration is limited by an upstream FEB at $\LfebUpS=-1000\,\rgz$, and a downstream FEB at $\LfebDS=+1000\,\rgz$.
\label{fig:PF_gam10}}
\end{figure}

\begin{figure}
\epsscale{1.00}
\plotone{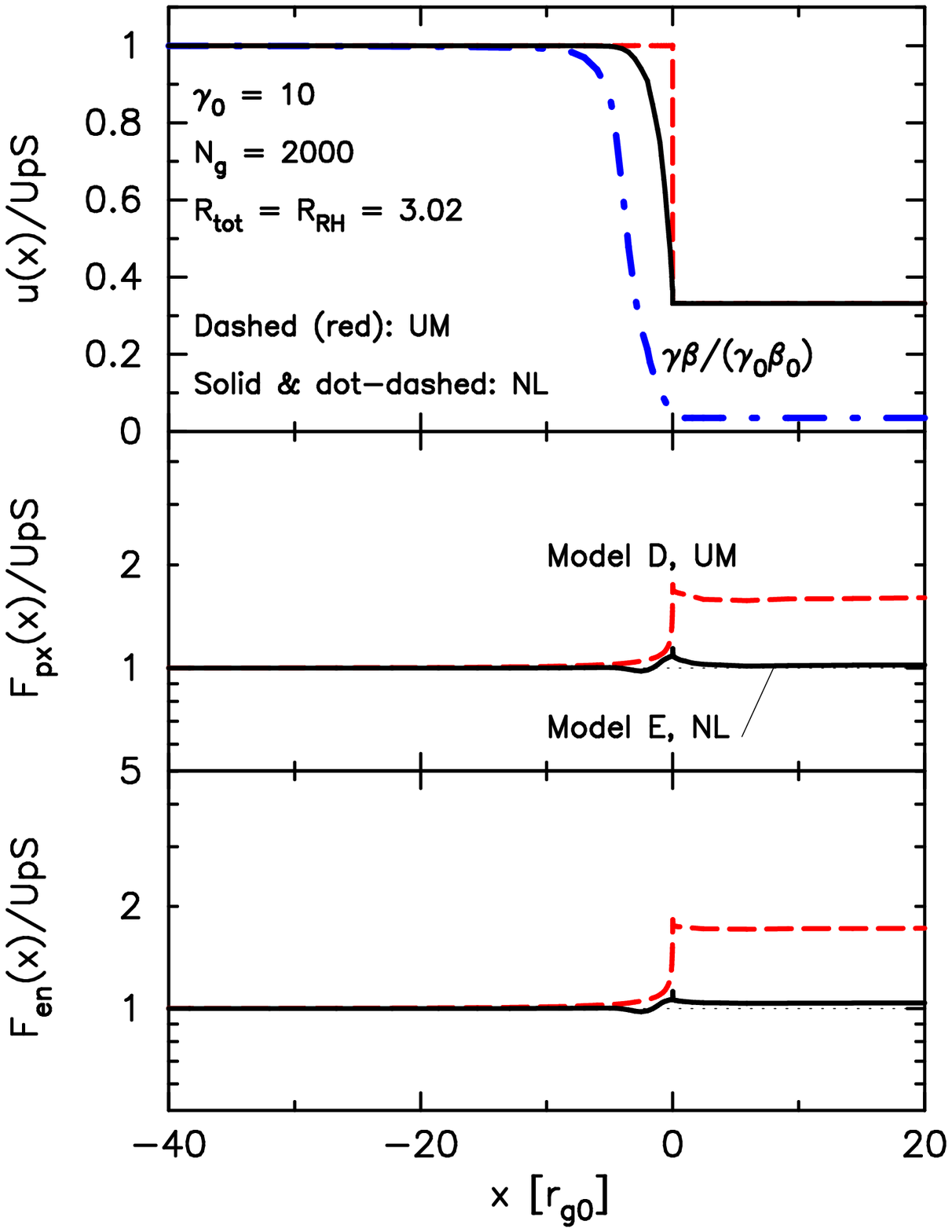}           % Fig 5
\caption{The top panel shows the shock structure for the \UM\ case 
(dashed red curve, Model \dDD) and the \NL\ case 
(solid black and dot-dashed blue curves, Model \eEE). The solid (black) and dashed (red) curves in the top panel are the flow speed $u(x)/u_0$ while the dot-dashed (blue) curve is $\GBeta/(\GBetaZ$),
where $\GBeta$ scales as 1/density 
for the \NL\ shock. The middle and bottom panels show the momentum and energy fluxes, respectively, normalized to far upstream values. Note that three particle species, protons, \HeT, and electrons, are included in determining the \SC\ shock structure.  
\label{fig:grid_gam10}}
\end{figure}

\section{Results}
We approximate the geometry of a \rel\ afterglow
shock moving in a jet as shown in 
Fig.~\ref{fig:Jet_cone}.
A detail of this situation is shown in Fig.~\ref{fig:Jet_detail}, where we show the additional approximations for this preliminary work that the shock is locally plane, the distances to the upstream or 
downstream FEBs ($\LfebUpS$ and $\LfebDS$) measured from the subshock are small enough so the jet cone is approximately a cylinder in the region surrounding the shock, and the diameter of the ``cylinder" is large enough so particle escape out the sides of the cone is negligible.
The cone material outside the upstream FEB is assumed to be stationary, i.e., it is in the local ISM or explosion frame.
Outside of the region between $L_\mathrm{UpS}$ and $L_\mathrm{DwS}$ we assume all accelerated particles have decoupled from the plasma and we ignore any emission they might produce.
A more realistic GRB afterglow model following the evolution of a jet shock is given in \citet{Warren2015dis}.

With these approximations  we 
calculate the shock structure and particle spectra for a given set of parameters, as listed in Table~\ref{tab:param}. First
we consider unmodified (UM) shocks where the backreaction of the accelerated particles on the shock structure is ignored.
We note an essential difference between unmodified shocks 
and \TP\ (TP) ones. A TP shock is one where the injection and acceleration efficiencies are low enough so the backpressure from accelerated particles can be ignored. In the TP case, to the limit of total energy placed in accelerated particles, momentum and energy can be conserved  without modifying the shock structure. Our thermal leakage injection model is efficient enough so TP shocks are never produced for the parameters we use here. In our UM examples, particles are injected and accelerated efficiently but the effect of shock accelerated particles on the shock structure is ignored. This provides a direct comparison to \NL\ (NL) shocks, where momentum and energy are conserved. Our main point in this paper is that, if the acceleration is efficient, the shock must be modified by the accelerated particles. 

While we do not 
consider TP shocks explicitly, the superthermal particle fluxes and photon emission from our UM examples can be simply re-scaled to a TP result. For example, if a TP result is defined as one where superthermal particles contain $\leq 1$\% of the total energy flux, so energy flux  will be conserved to within 1\%, our UM superthermal particle and photon emission needs to be reduced by a factor $\Ntp \geq  100 \times [\Fen(x>0)/\FenZ]$, where $\FenZ$ is the far upstream energy flux and $\Fen(x>0)$ is the UM downstream  energy flux. For the UM $\gamZ=10$ examples we show below, $\Fen(x>0)/\FenZ \sim  2$.

\subsection{Unmodifed (UM) Examples}
In Fig.~\ref{fig:UM_gam10} we show number spectra, $dN/dp$, calculated just downstream from an unmodified  shock with $\gamZ=10$. 
The solid and dotted black curves are protons and the red curves are electrons.
The dotted curves were calculated 
for $\EnTran=0$, while the solid curves were calculated with $\EnTran=0.15$, i.e., 15\% of the proton ram kinetic energy is transferred to electrons as particles first cross $x=0$ headed downstream.\footnote{We refer to the sharp drop in $u(x)$ that occurs at $x\simeq 0$ as the subshock 
(see Figs.~\ref{fig:grid_gam10} and \ref{fig:grid_gam2}). For an \UM\ shock, there is no distinction between the shock and subshock.}
Since all particles injected far upstream are \nonrel, 
the transformation from upstream to downstream frames strongly favors more massive particles in the first shock crossing. This results in the strong depression of electrons relative to \Hone. Once all particles become \rel\ they are treated equally and obtain similar power laws, i.e., the UM shocks in Fig.~\ref{fig:UM_gam10} show the canonical $dN/dp \propto p^{-2.23}$ power law above the ``thermal" peak and below the high momentum cutoff.
For the protons, the cutoff  at $\sim 10^7\, m_pc$ is produced by an upstream FEB at $\LfebUpS=-1\xx{4}\,\rgz$, where $\rgz=\etamfp m_p u_0 c/(e B_0)$. 
The electrons cut off at a lower momentum ($\sim 2\xx{6}\, m_pc$) due to radiation losses. Without radiation losses, the electrons would obtain the same $\pmax$ as protons since $\pmax$ scales as $Z$.
The normalization of the electron spectra shows the dramatic effect of $\EnTran$: the $e/p$ ratio  is increased by nearly three orders of magnitude with $\EnTran=0.15$. The proton normalization is only slightly influenced by $\EnTran$. 

To save computation time, the models in Fig.~\ref{fig:UM_gam10} used a 
probability-of-return calculation instead of a downstream FEB.  This mimics an infinite downstream region and allows rapid acceleration to high energies
\citep[see][ for a discussion of the probability of return calculation]{EBJ96}.
In Fig.~\ref{fig:PF_gam10} we show spectra from an UM shock  measured  in the local plasma frame with both upstream and  downstream FEBs. While both FEBs are present, typically one (the shorter, measured in diffusion lengths) 
will determine the maximum momentum $\pmax$.
For Fig.~\ref{fig:PF_gam10}, both $\LfebUpS$ and $\LfebDS$ are $1\xx{3}\,\rgz$ from the subshock, but $\LfebDS$ determines $\pmax$ since it is much more difficult for particles to stream away from the shock in the upstream region with $\gamZ=10$.

The FEBs also determine the total number of particles 
accelerated (as indicated in Fig.~\ref{fig:Jet_detail}) 
and the $dN/dp$ spectra in 
Fig.~\ref{fig:PF_gam10} are normalized to the total particle number in a region surrounding an observation position $x$ (the position is indicated in Fig.~\ref{fig:PF_gam10}). 
While the particular number of accelerated particles in these examples is arbitrary,\footnote{The normalization of the spectra in 
Fig.~\ref{fig:PF_gam10} 
(and Figs.~\ref{fig:PF_NL_gam10} and \ref{fig:gam1p5_dNdp} below) depends on the opening angle and length of the jet, the ambient number density, and the positions of the FEBs 
\citep[see][ for a full discussion]{Warren2015dis}.} 
it is important to note that the \mc\ code determines the absolute number of accelerated particles, and subsequent radiation, for any given set of environmental and shock parameters, including $\EnTran$.
As long as $\gamZ \gg 1$, the large majority of accelerated particles will be in the downstream region since it is difficult for particles to stream upstream into the shock precursor. This is reflected in the higher normalizations of the black curves, measured at $x=0+$ (i.e., downstream), compared to  the red or blue curves, measured in the shock precursor at $x=-1\,\rgz$ and $x=-120\,\rgz$, respectively.

\begin{figure}
\epsscale{1.00}
\plotone{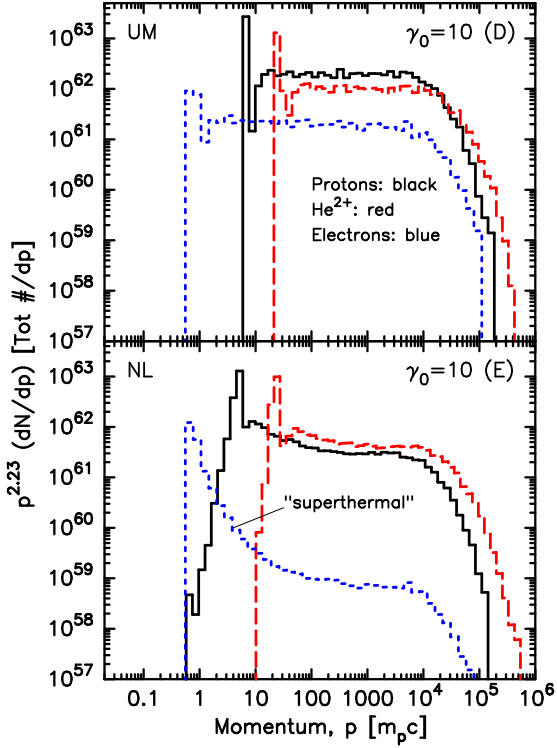}           % Fig 6   
\caption{Downstream, local plasma frame (LPF) spectra for the \UM\ shock shown in Fig.~\ref{fig:grid_gam10} (top panel, Model \dDD) and the \NL\ shock shown in Fig.~\ref{fig:grid_gam10} (bottom panel, 
Model \eEE). Note the pronounced ``superthermal" tail on the electron distribution.
\label{fig:PF_NL_gam10}}
\end{figure}

\begin{figure}
\epsscale{1.00}
\plotone{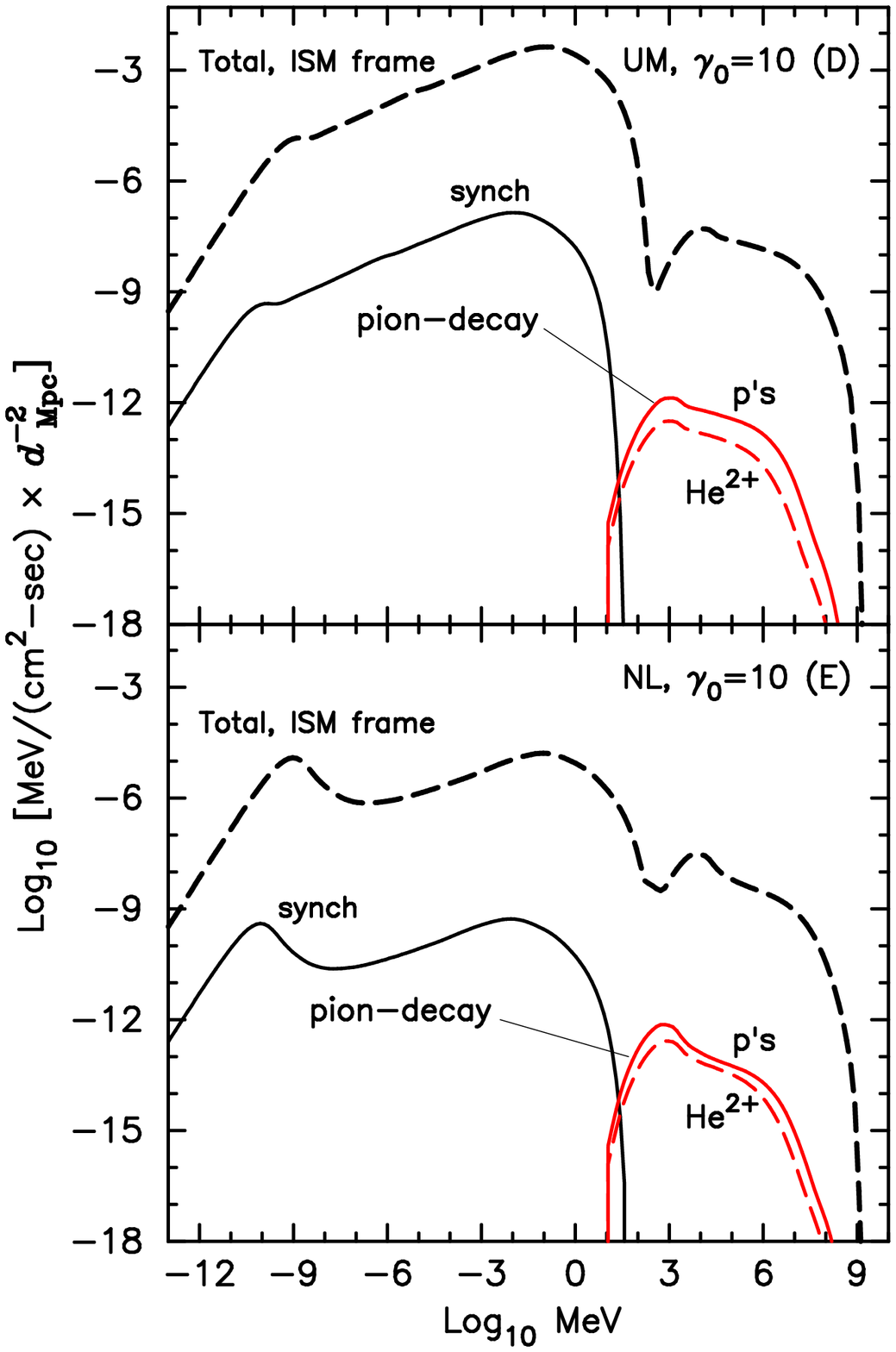}           % Fig 7 
\caption{Photon emission for the UM (Model \dDD) and NL (Model \eEE)
shocks shown in 
Figs.~\ref{fig:grid_gam10} and \ref{fig:PF_NL_gam10}. 
The dashed (black) curves are the total emission transformed to the ISM frame for an observer at $\Dist = -\dMpc$\,Mpc, at an angle within $1/\gamZ$ from the shock normal. The lower \synch\ and \pion\ curves in each panel show emission calculated in the local plasma frame and summed over the shock from $\LfebUpS=-300\,\rgz$ to $\LfebDS=+1000\,\rgz$.
The dashed (red) \pion\ curves are the LPF emission from \HeT. 
\label{fig:phot_gam10}}
\end{figure}

Besides the downstream FEB and normalization, 
the spectra in Fig.~\ref{fig:PF_gam10} 
differ from those in Fig.~\ref{fig:UM_gam10} in that they are shown in the local plasma frame and they include spectra calculated upstream from the subshock (red and blue curves), as well as downstream (black curves).
The effect of the \LorT\ from the shock to the local plasma frame is clearly indicated by the high-momentum upstream spectra (red and blue curves) which extend to higher momentum than the downstream spectra. Relative to the subshock, the upstream plasma frame moves  with $\gamZ=10$, while the downstream plasma frame moves with 
$\gamma_2=[1-(u_2/c)^2]^{-1/2} \simeq 1.06$. Here $u_2$ is the downstream bulk plasma speed measured in the shock frame.

\subsection{Nonlinear Examples, $\gamZ=10$}
In Fig.~\ref{fig:grid_gam10}  we show the structure of a shock where the backpressure from accelerated protons, \HeT, and electrons is taken into account. For this example, $\EnTran=0.1$, and the  electron  pressure contributes to the determination of the \SC\ shock structure. Here
$\LfebUpS=-300\,\rgz$ and $\LfebDS=+1000\,\rgz$.  
In the top panels, the solid (black) curve is the bulk flow speed, $u(x)/u_0$, and the dot-dashed (blue) curve is $\GBeta/(\GBetaZ)$, where $\betaX=u(x)/c$, $\gamX=[1 - \betaX^2]^{-1/2}$, and $\betaZ=u_0/c$.
The dashed (red) curve is $u(x)/u_0$ for the UM case. The lower panels show the momentum and energy fluxes for the NL case (solid black curves), as well as for the UM case (dashed red curves). All curves are normalized to far upstream values. Without shock smoothing, the downstream momentum and energy fluxes are nearly a factor of two out of conservation.

We note that in \nonrel\ and \transrel\ shocks, the smoothing required for momentum and energy conservation is accompanied by an increase in the overall shock compression ratio above the \RH\ value, i.e., $\Rtot > \rRH$ \citep[e.g.,][]{BE99}. The overall compression ratio is defined as $\Rtot=u_0/u_2$ and the subshock compression ratio is $\Rsub=u_1/u_2$. Here, $u_2$ is the bulk plasma speed downstream from the subshock and $u_1$ is the plasma speed just upstream of the viscous subshock, both measured in the shock rest frame.\footnote{As seen in Fig.~\ref{fig:grid_gam10}, the definition of $u_1$ is imprecise because the \mc\ solution allows for a smooth decrease in the precursor speed into the downstream region.  
Part of the increase in $\Rtot$ is due to escaping particles acting to soften the equation of state for the plasma that remains coupled to the shock system.  The remainder of the increase is caused by the ratio of specific heats for particles downstream of the shock: as the shock speed increases, the particles receive more energy from the first shock crossing, the average particle approaches relativistic energies, and the ratio of specific heats drops from
$5/3 \rightarrow 4/3$. 
For fully \rel\ shocks, however, $\Rtot \simeq \rRH$ \citep[][]{DoubleEtal2004}; there is minimal particle escape, 
and the downstream ratio of specific heats is already $\sim 4/3$ and so cannot decrease further. 
The compression ratio in Fig.~\ref{fig:grid_gam10} reflects this invariance, as it is the same in the unmodified and nonlinear cases.}

The top panel in Fig.~\ref{fig:PF_NL_gam10} shows downstream proton, \HeT, and electron spectra for the UM shock
(dashed curves in Fig.~\ref{fig:grid_gam10}). The bottom panel shows these spectra from the \NL\ shock. All parameters are the same for these two cases---the only difference is that  momentum and energy are conserved in the NL case.

The effects of the smooth shock structure are clearly evident. 
In the NL case, the downstream spectra are noticeably curved and less intense, as is necessary to conserve energy and  momentum. 
Significantly, the electrons are more modified than the protons or \HeT\ and the $e/p$ ratio, in the quasi-power law portion of the spectra, drops by more than an order of magnitude. With $\EnTran=0$, $e/p$ would have dropped by several more orders.
This difference is a direct result of our scattering 
assumption, i.e., equation~(\ref{eq:mfp_two}). The electrons, with their small \AZ, feel the effects of the smooth shock more acutely than the heavier ions and are less efficiently injected and accelerated until they reach 
$p \gsim 10\, m_p c$. 
In sources where electrons are assumed to have a large fraction of the energy budget,  the efficiency for accelerating electrons cannot be orders of magnitude less than it is for protons. Therefore energy must be transferred from heavy particles to electrons with a reasonable
efficiency if  \Facc\ is to be important 
\citep[e.g.,][]{SironiSpit2011,SSA2013}. 

The heavier \HeT, with $A/Z = 2$, is accelerated more efficiently than protons; the $\HeP$ ratio above $\sim 100\,m_p c$ goes from $\HeP < 1$ in the UM shock to $\HeP > 1$ in the NL shock (see Fig.~\ref{fig:dndp_AZ}). This is particularly significant at the high momentum cutoff. The \AZ\ effect is discussed in more detail in Section~\ref{sec:AZ}.
For Model \eEE\ in Fig.~\ref{fig:PF_NL_gam10}, the fraction of total ram kinetic energy placed in particles of 100\,MeV or greater, is $\EffGeVp=0.60$, $\EffGeVhe=0.30$, and $\EffGeVe=0.10$, for protons, \HeT, and electrons, respectively (see Table~\ref{tab:param} where it is noted that these fractions are measured in the shock frame.)

The effects of shock smoothing also show up in the broadening and shift to lower momentum of the ``thermal" peaks, as seen in the bottom panel of 
Fig.~\ref{fig:PF_NL_gam10}. 
Of particular interest is the pronounced ``superthermal" tail the electrons obtain in the NL case.
If \SSA\ (SSA) is unimportant, this can produce a notable effect in the \synch\ emission, as we discuss next.

\subsection{Photon Emission, $\gamZ=10$}
In Fig.~\ref{fig:phot_gam10} we show the photon emission 
produced by the shocks described in Figs.~\ref{fig:grid_gam10} and \ref{fig:PF_NL_gam10}. As in Fig.~\ref{fig:PF_NL_gam10}, the top panel is for the \UM\ shock and the bottom panel is for the \NL\ shock---all other parameters are the same. 
The curves labeled \synch\ and \pion\ show isotropic emission calculated in the local plasma frame (LPF) summed over the regions between the upstream and downstream FEBs (i.e., between $x=-300\,\rgz$ and $x=+1000\,\rgz$) as indicated in 
Fig.~\ref{fig:Jet_detail}.
These are the fluxes that would be observed at a distance $\dMpc$\,Mpc if no \Lor\ transformations were required. Here, $\dMpc$ is the distance in Mpc. 
Of course the particle distributions will not be isotropic in the ISM frame, and \Lor\ transformations are required. The dashed (black) curves  show the total emission from \synch, IC, and \pion\ transformed to the ISM frame  and seen by an observer at $\Dist =-\dMpc$\,Mpc within an angle $1/\gamZ$ from the shock normal.

The shock simulation is done in the shock rest frame. To obtain the \LPF\ \synch\ and \pion\ spectra we first transform the particle spectra to the LPF taking into account the anisotropies introduced by the \rel\ flow.  
We then calculate the photon emission in the LPF assuming it is produced isotropically. This is reasonable for the \synch\ emission since we implicitly assume the background magnetic field is highly turbulent, and the \synch\ photons should be produced isotropically as the electrons spiral in the turbulent field. It is also a good approximation for the \pion\ emission since the protons and \HeT\ ions interact with the local plasma and  produce pions that can  become isotropic in the LPF before emitting a \gamray.

From this isotropic emission, we obtain the flux in the ISM frame  ahead of the shock 
by employing the standard Doppler shift and \Lor\ transformations \citep[e.g.,][]{KumarZ2014}. 
For an observer at $\Dist = -\dMpc$\,Mpc, within an angle $\Aobs =1/\gamZ$ from the jet direction,
these give a boost to the energy flux $\propto \Dop^4$, where $\Dop = \gamma(x) [1 + \beta(x)\cos{\Aobs}]$ is the Doppler 
factor at position $x$ relative to the subshock.
One factor of $\Dop$ comes from the Doppler shift, one from time dilation, and two from the \rel\ beaming. 
Here $\gamma(x)$ and $\beta(x)$ are measured relative to the ISM frame and will be different for each region between the upstream and downstream FEBs as shown in Fig.~\ref{fig:Jet_detail}. Far upstream $\Dop=1$, while downstream for $\gamZ=10$ and $\Rtot \simeq 3.02$, $\Dop\simeq 14$. 

Since the \CMB\ (CMB) photons are nearly isotropic in the ISM frame, we calculate the IC emission in a different fashion.\footnote{We only consider CMB photons here for simplicity---the techniques we present can be generalized to include IC emission from other photon fields including synchrotron self-Compton photons if the jet parameters warrant it.} 
We first transform the electron distribution into the ISM frame keeping the two-dimensional anisotropy inherent in our plane-parallel shock simulation. We then calculate the emission directed toward the observer at $\Dist =-\dMpc$\,Mpc using equation~(9) in \citet{Jones68}. The assumption here is that the \rel\ electrons produce strongly beamed emission, so only those electrons directed toward the observer contribute to the observed flux.
In Fig.~\ref{fig:phot_gam10}, the dashed (black) curves contain the full observed flux from  emission produced over the modified shock structure and transformed to the ISM frame for an observer within $1/\gamZ$ from the jet axis 
(i.e., toward $+x$ in Fig.~\ref{fig:Jet_detail}). The IC emission is part of this sum.

The effects from shock smoothing on the particle distributions 
(Fig.~\ref{fig:PF_NL_gam10}) produce corresponding changes in the photon emission. Since electrons are suppressed more than ions in the NL shock, the \synch\ and IC emission drops more than \pion\ between the UM and NL cases. In contrast, the \pion\ from \HeT\ (dashed red \pion\ curve) is increased relative to that from protons by the \AZ\ effect.
The ``thermal" peaks  near 1\,$m_pc$ for electrons, and near 10\,$m_pc$ for protons and \HeT, show up as clear peaks in the 
\synch\ and 
\pion\ emission 
at $E \sim 10^{-11}$ and $10^{3}$\,MeV, respectively. 
This is particularly significant for the \synch\ emission near $10^{-9}$\,MeV.
The NL curvature in the particle spectra shows clearly in the individual components and remains strongly
evident in the summed flux (bottom panel of Fig.~\ref{fig:phot_gam10}).

\begin{figure}
\epsscale{1.00}
\plotone{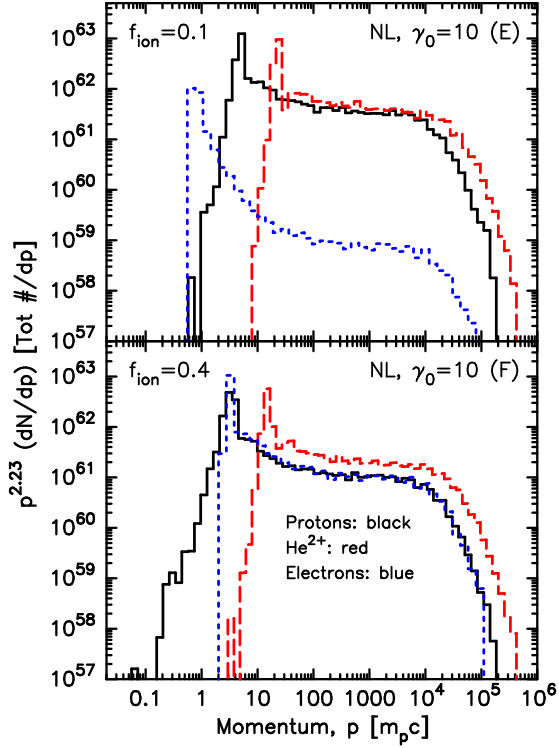}           % Fig 8  
\caption{Nonlinear downstream LPF spectra for 
Model \eEE\ ($\EnTran=0.1$) and Model \fFF\ ($\EnTran=0.4$). In both panels the solid (black) curves are protons, the dashed (red) curves are \HeT, and the dotted (blue) curves are electrons.
\label{fig:dNdp_fion}}
\end{figure}

\begin{figure}
\epsscale{1.00}
\plotone{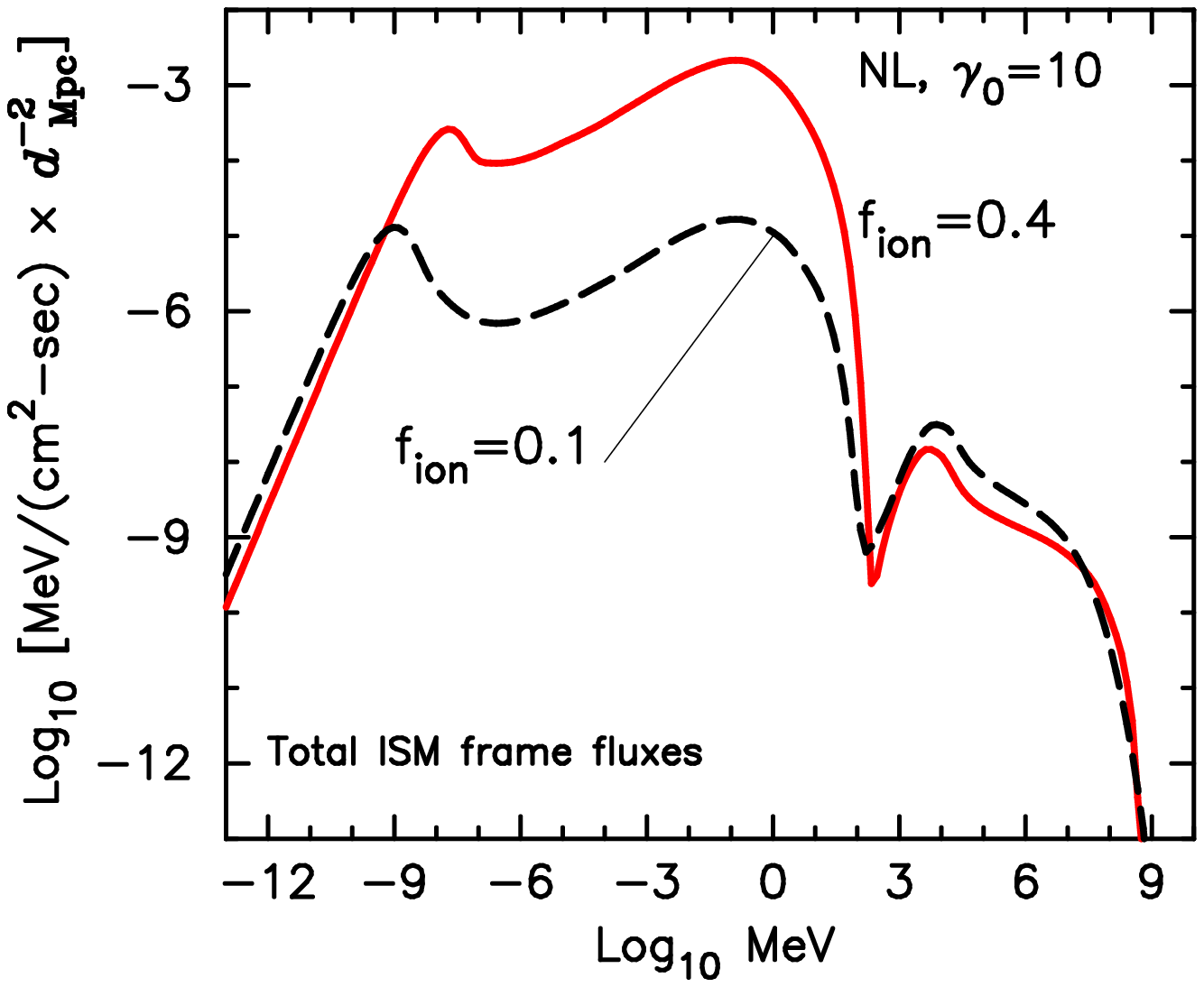}           % Fig 9  
\caption{Total observed energy flux for NL Models \eEE\ ($\EnTran=0.1$) and \fFF\ ($\EnTran=0.4$).
\label{fig:phot_fion}}
\end{figure}

\begin{figure}
\epsscale{1.00}
\plotone{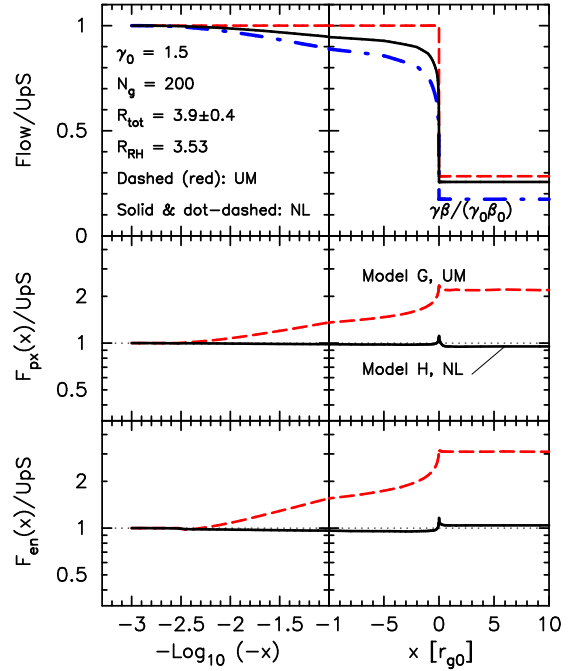}           % Fig 10  
\caption{All curves are as in Fig.~\ref{fig:grid_gam10} for $\gamZ=1.5$. The dashed (red) curves are the UM (Model \gGG) case while the solid (black) and dot-dashed (blue) curves are the 
NL (Model \hHH) profiles. The upstream and downstream FEBs are $\LfebUpS =-300\,\rgz $ and $\LfebDS =+1000\,\rgz$, as in Fig.~\ref{fig:grid_gam10}, but these distances differ in absolute units since $u_0$ varies between $\gamZ=10$ and $\gamZ=1.5$.   
\label{fig:grid_gam2}}
\end{figure}

\begin{figure}
\epsscale{1.00}
\plotone{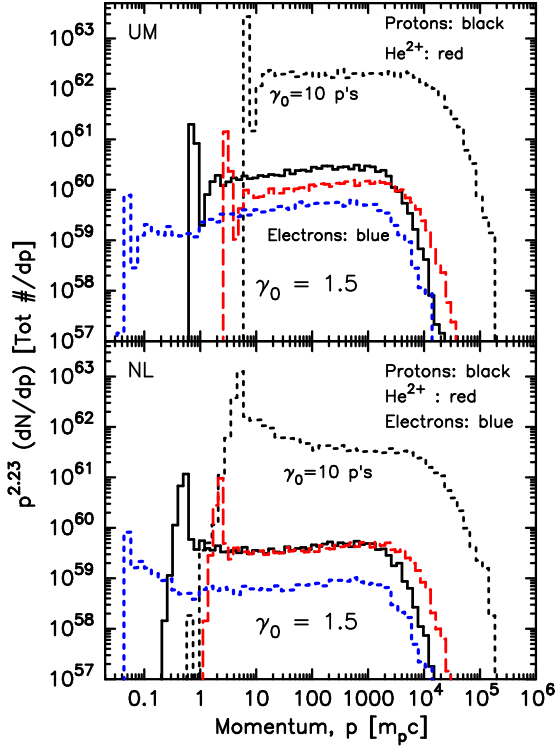}           % Fig 11  
\caption{Except for the dotted (black) 
curves  labeled `$\gamZ=10$ p's' all spectra are downstream, local plasma frame (LPF) spectra for the \UM\ $\gamZ=1.5$ shock (top panel, Model \gGG) 
and the \NL\ $\gamZ=1.5$ shock (bottom panel, Model \hHH) 
shown in Fig.~\ref{fig:grid_gam2}. 
The curves labeled `$\gamZ=10$ p's'  are identical to those in 
Fig.~\ref{fig:PF_NL_gam10}.
\label{fig:gam1p5_dNdp}}
\end{figure}

\begin{figure}
\epsscale{1.00}
\plotone{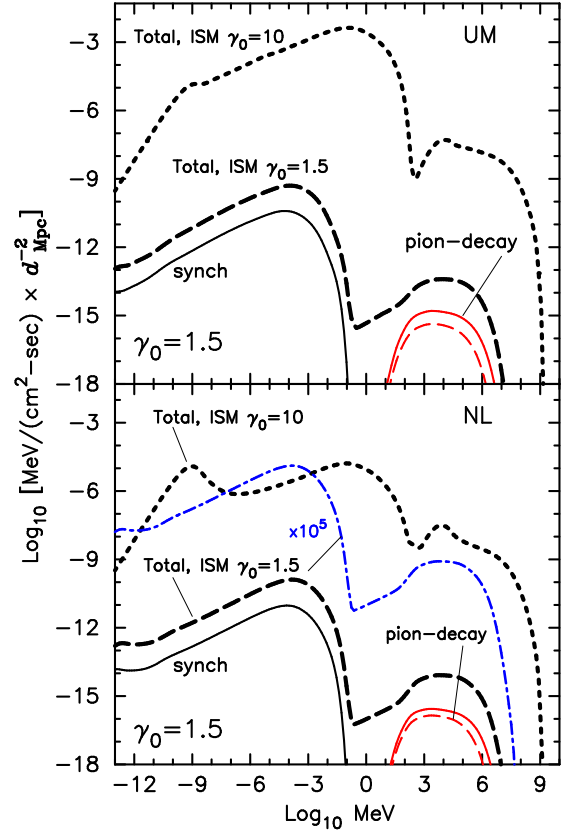}           % Fig 12  
\caption{Photon emission for the UM (Model \gGG) and NL (Model \hHH) shocks shown in Figs.~\ref{fig:grid_gam2} and \ref{fig:gam1p5_dNdp}. 
Except for the dotted (black) 
curves  labeled `Total, ISM $\gamZ=10$', all curves are for the $\gamZ=1.5$ shock and are in the same format as in Fig.~\ref{fig:phot_gam10}.
The dashed (black) curves are the total emission transformed to the ISM frame for an observer at $x = -\dMpc$\,Mpc, at an angle within $1/\gamZ$ from the shock normal. 
The dot-dashed (blue) curve in the lower panel is the total ISM frame $\gamZ=1.5$ emission multiplied by $10^5$.
\label{fig:phot_gam1p5}}
\end{figure}

\begin{figure}
\epsscale{1.00}
\plotone{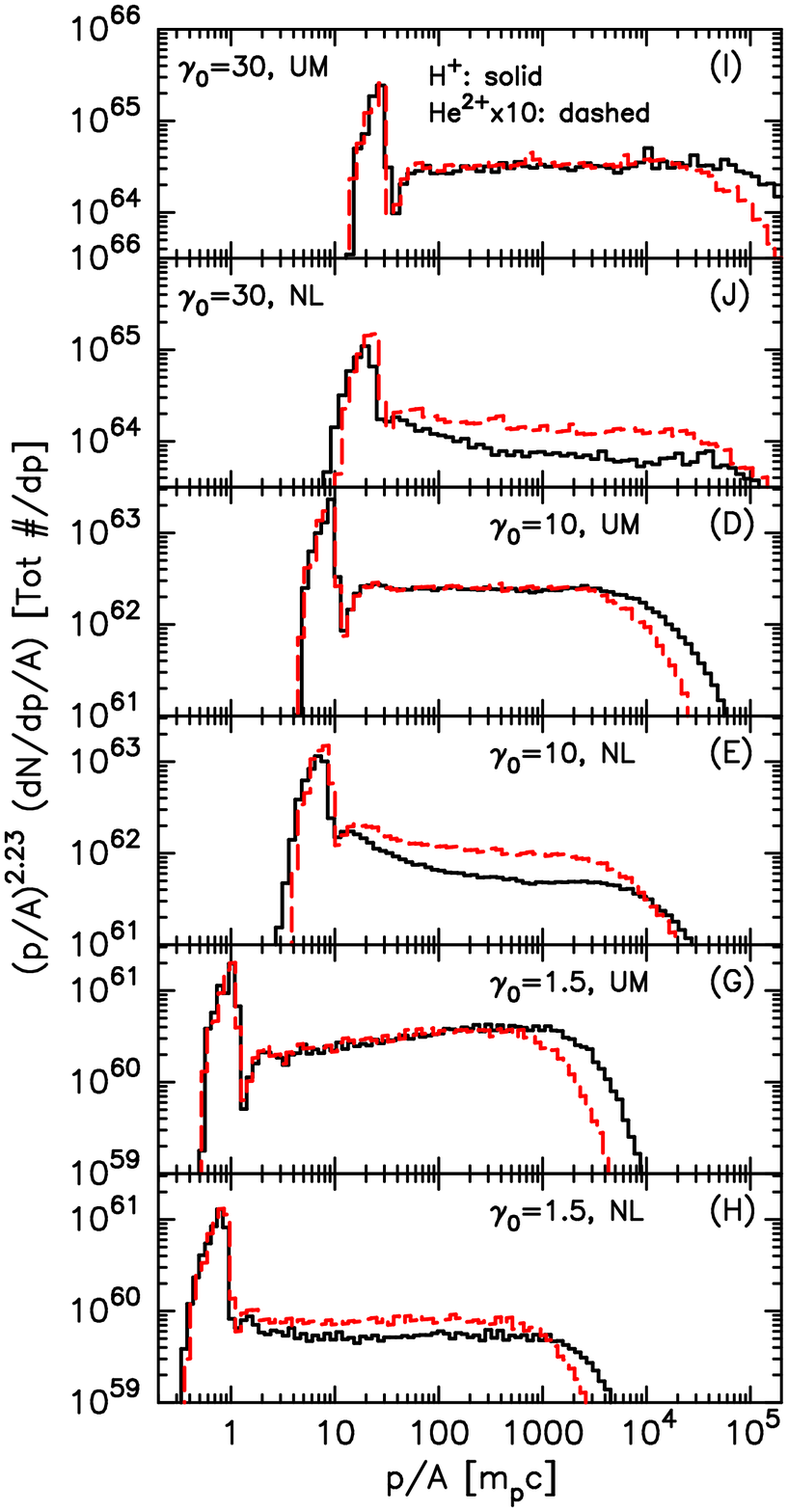}           % Fig 13  
\caption{Models \iII, \dDD, and \gGG\ show proton and \HeT\ spectra for UM shocks, while models \jJJ, \eEE, and \hHH\ are the corresponding spectra for the NL shocks. In all cases, the \HeT\ spectra are multiplied by 10 to adjust for the ambient number density. When plotted in $p/A$ units the UM spectra are identical except for statistics and the maximum momentum cutoff.  The NL shocks show a clear \AZ\ enhancement in the \HeTp\ ratio.
\label{fig:dndp_AZ}}
\end{figure}

\
\subsection{Nonlinear, $\gamZ=10$, Variation of $\EnTran$}
While we have used $\EnTran=0.1$ in our NL Model \eEE, the PIC simulations of \citet{SSA2013} show examples where $\sim 40\%$ of the energy in accelerated particles ends up in electrons  
\citep[see figure~11 in][]{SSA2013}.
In Fig.~\ref{fig:dNdp_fion} we compare particle spectra, and in
Fig.~\ref{fig:phot_fion} the total observed energy flux, for $\EnTran=0.1$ (Model \eEE) and $\EnTran=0.4$ (Model \fFF). The flux between $\sim 10^{-9}$\,MeV and  $\sim 1$\,GeV is $\sim 100$ times greater for the $\EnTran=0.4$ case, with a much smaller decrease in the GeV-TeV emission. 
The $\EnTran=0.4$ example shows the curved spectral shape that results from the NL shock smoothing but it is less pronounced 
between $10^{-9}$ and 1\,MeV than for $\EnTran=0.1$. In the GeV-TeV range, the curvature is slightly greater than for $\EnTran=0.4$.  

With $\EnTran=0.4$, the energy distribution above 100\,MeV in protons, \HeT, and electrons is 
$\EffGeVp=0.36$, $\EffGeVhe=0.20$, and $\EffGeVe=0.44$, respectively; nearly 50\% of the ram kinetic energy goes into 100\,MeV or greater electrons, as measured in the shock frame.
The large difference in electron normalization between the $\EnTran=0.1$ and $0.4$ cases comes about mainly from the NL shock smoothing effects the electrons with $\EnTran=0.1$ receive. When $\EnTran=0.4$, the lowest energy downstream electrons that cross back upstream have a long enough diffusion length so they feel a large effective compression ratio. Nonlinear effects reduce the electron acceleration more for $\EnTran=0.1$ than for $\EnTran =0.4$.

\subsection{\Transrel, $\gamZ=1.5$}
As described in \citet{EWB2013}, the \mc\ simulation smoothly treats \nonrel\ to \ultrarel\ shocks. 
In Fig.~\ref{fig:grid_gam2} we show the profile of a \transrel\ $\gamZ=1.5$ shock for comparison with Fig.~\ref{fig:grid_gam10}. Apart from $\gamZ$, all input parameters are the same for the $\gamZ=1.5$ and $\gamZ=10$ cases.
Since it is easier for accelerated particles to diffuse upstream against the inflowing plasma with $\gamZ=1.5$ than against $\gamZ=10$, the NL shock precursor is much more extended than it is for $\gamZ=10$. The bulk flow speed, $u(x)$, is noticeably modified out to 
$x=\LfebUpS=-300\,\rgz$ (note the split log-linear $x$-axis in 
Fig.~\ref{fig:grid_gam2}). 

As in the $\gamZ=10$ case, the momentum and energy fluxes are not conserved in the UM shock but are within a few percent of the far upstream values
once the shock structure is modified by the CR backpressure. For $\gamZ=1.5$, the overall compression ratio must also be increased to conserve momentum and energy and we find $\Rtot = 3.9\pm 0.4$, where the uncertainty comes from statistics and errors inherent in the \mc\ smoothing algorithm. This result is similar to that given in 
\citet{EWB2013} except here we have included \HeT\ and electrons in determining the \SC\ shock structure, and we use a downstream FEB as well as an upstream one.

In Fig.~\ref{fig:gam1p5_dNdp} we show the particle spectra for $\gamZ=1.5$ in the same format as Fig.~\ref{fig:PF_NL_gam10} except we have added the $\gamZ=10$ proton spectra for comparison. The $\gamZ=1.5$ spectra are harder than those for $\gamZ=10$ mainly because $\Rtot$ is larger.
A comparison of the $e/p$ ratio at $\sim 300\, m_p c$ for the NL models in Fig.~\ref{fig:PF_NL_gam10} ($e/p \sim 1/300$)  and 
Fig.~\ref{fig:gam1p5_dNdp} ($e/p \sim 1/6$), shows that the $\gamZ=1.5$ shock is much more effective in injecting and accelerating electrons than the $\gamZ=10$ shock.
However, the maximum momentum is noticeably lower for $\gamZ=1.5$ and the ``thermal" peak is also lower since, for downstream spectra, it occurs at $\sim \gamZ m_p c$ for protons in the UM shock. The ``thermal" peak is at a noticeably lower momentum in the NL case.  

Another important difference is the normalization. The $\gamZ=10$ proton distributions are about a factor of 100 above the $\gamZ=1.5$ spectra in the quasi-power law region. This comes about for two primary reasons. The 
$\gamZ=10$ shock has a considerably higher downstream density since
$n_2 = \gamZ \betaZ n_0/(\gamT \betaT)$
and, with $\rgz$ being larger, the number of particles in the distribution is larger. Another cause is that the spectra are multiplied by $p^{2.23}$ rather than by $p$.
The area under a $dN/d\log(p)$ curve would be the total number of particles.  But $dN/d\log(p) = p \cdot dN/dp$, so the extra 1.23 powers of $p$ merely enhance the perceived importance of particles at higher momentum. Since the $\gamZ = 10$ shock contains particles at higher momenta, it also gets plotted higher on the $p^{2.23} dN/dp$ plot.
The efficiencies for producing 100\,MeV or greater energy particles are $\EffGeVp=0.52$, $\EffGeVhe=0.27$, and $\EffGeVe=0.08$ for Model \hHH; i.e., the $\gamZ=1.5$ shock puts $\sim 8$\% of the shock energy into energetic electrons.

In Fig.~\ref{fig:phot_gam1p5} we show the photon emission for $\gamZ=1.5$ in the same format as Fig.~\ref{fig:phot_gam10}, with the addition of the total ISM frame emission for the $\gamZ=10$ 
cases (dotted, black curves) added to the $\gamZ=1.5$ cases (dashed, black curves) for comparison. The total emission for an observer at 
$\Dist =-\dMpc$\,Mpc is dramatically different for the two \Lor\ factors. 
To highlight this in the lower panel, we have plotted the total ISM emission for $\gamZ=1.5$ multiplied by $10^5$ 
(dot-dashed, purple curve). The broadband spectral shapes are very different between $\gamZ=10$ and $1.5$, and at $\sim 1$\,keV the normalization differs by $\sim 10^5$.

Note that the NL $\gamZ=1.5$ spectrum is harder than $\gamZ=10$ in the GeV-TeV range but cuts off at a lower energy. The lower cutoff energy shows up dramatically in the \synch\ emission. For $\gamZ=1.5$ the \synch\ peak is around 1\,keV, typical of SNR observations, while for $\gamZ=10$, the peak is around 1\,MeV.
At radio emitting energies, the \synch\ spectra are very different because of the emission produced by the downstream thermal electrons. 
Of course we have not considered SSA here and this process will produce a low-energy cutoff in the \synch\ emission which may mask the emission from the thermal electrons. Since the frequencies of the thermal peak and SSA cutoff depend on $B$ as   $\nuPk \propto B$ and $\nuSSA \propto B^{1/5}$, respectively, a stronger magnetic field makes it more likely that the thermal peak will be resolved. 
We have also not considered \gamray\ absorption between the GRB and Earth.

\subsection{$A/Z$ Enhancement of Heavy Ions} \label{sec:AZ}
The \mc\ code assumes that all scatterings are elastic in the local plasma frame. 
This implies that an insignificant fraction of the particle energy is transferred to magnetic turbulence in the wave generation process. 
With elastic scattering, $\gamma_i v_i \propto p_i/A$ remains constant in a scattering event, where $\gamma_i$ and $ v_i$ are local frame values. The  energy gain particles receive on crossing the shock, which is determined by a \Lor\ transformation between the two frames, also scales as $p_i/A$. 
In our plane-parallel approximation,the probability that particles make a set number of shock crossings also depends only on $\gamma_i v_i$.
Thus an UM shock, with the \mc\ assumptions, will treat all particles identically in momentum per nucleon, including the thermal leakage injection.

An exception to this occurs if the acceleration is limited by a boundary at a fixed distance, as we assume here, or by a maximum acceleration time. Since we assume 
equation~(\ref{eq:mfp_two}), diffusion length and acceleration time 
both scale as $(A/Z) (p_i/A)$ for $v_i \sim c$. 
Thus, apart from the  normalization set by input parameters and the maximum momentum cutoff, all species should have identical spectra when plotted against $p/A$. With a fixed FEB, high \AZ\ particles will turn over at a lower $p/A$ than low \AZ\ particles. 

For models \iII, \dDD, and \gGG\ in Fig.~\ref{fig:dndp_AZ}, we show downstream, shock frame, proton and \HeT\ spectra for UM shocks plotted in $p/A$ units. Except for statistical variations, a factor of 10 normalization since $n_\alpha = 0.1 n_p$, and the high momentum cutoff, the proton and \HeT\ spectra are identical. Electrons are not plotted but would show the same effect. 
In the corresponding NL models \jJJ, \eEE, and \hHH, a clear enhancement of \HeT, produced solely 
because the shock structure in smoothed by the backpressure of accelerated particles, is seen.
For the $\gamZ=30$ Model \jJJ, $\EffGeVp=0.66$ and $\EffGeVhe=0.34$.

\newlistroman

The factor of two enhancement in the \HeTp\   ratio seen in the NL models in Fig.~\ref{fig:dndp_AZ} should be large enough to see clearly in PIC simulations. 
Since this enhancement is a prediction that stems directly from important  assumptions of efficient \Facc\ and thermal leakage injection, adding helium to PIC simulations can test these assumptions.
If the acceleration is efficient, and the \AZ\ effect is {\it not} seen, it implies that one or more of the following may be happening.  
\listromanDE
The accelerated protons and \HeT\ may be sharing significant energy with each other rather than interacting mainly elastically with the background turbulence. 
If this is the case, it will influence all aspects of \Facc.
\listromanDE
The particle mean free path may not be a monotonically increasing function of momentum or it may differ substantially for protons and \HeT.
\listromanDE
Different \AZ\ particles may interact differently with the viscous subshock layer.
A basic assumption for thermal leakage injection is that the subshock is essentially transparent, i.e., phenomena such as cross-shock potentials or large-scale turbulence do not strongly influence the injection process.
If these phenomena are important, it is likely they will influence  different \AZ\ particles differently, modifying the \AZ\ enhancement seen in Fig.~\ref{fig:dndp_AZ}.

\section{Conclusions} \label{sec:Conc}
As complicated as particle acceleration in \rel\ shocks may be, one aspect is profoundly simple: if the acceleration is efficient and a sizable fraction of the bulk plasma flow energy is put into individual accelerated particles, as is often assumed in applications
\citep[e.g.,][]{Kulkarni1999,Piran2001,Piran2004_05,Meszaros2006},  the accelerated particles must \SCly\ modify the shock structure to conserve momentum and energy regardless of the plasma physics details.
Assuming first-order Fermi acceleration, we have investigated how the kinematics of shock modification influences the relative acceleration of electrons, protons, and heavy elements (i.e., \HeT) using a \mc\ simulation with a dynamic range large enough to model acceleration from injection at \nonrel\ thermal energies to \ultrarel\ CR energies. Fig.~\ref{fig:PF_gam10} shows a 12 decade range in plasma-frame momentum and a greater than 20 decade range in $dN/dp$. A corresponding range in photon emission is also obtained 
(e.g., Fig.~\ref{fig:phot_gam1p5}).

The underlying wave-particle plasma interactions, which are parameterized in the \mc\ code, will influence details of the shock modification and the resultant radiation; they will determine if acceleration is, in fact, efficient and  set the maximum energy particles obtain.
However, our results show general aspects that are largely independent of the poorly known plasma physics details if the acceleration is efficient.
Considering only the kinematics, electrons will be accelerated much less efficiently than ions if the shock structure is modified by the heavy particles. This result assumes that the heavy particles and electrons diffuse in a similar fashion, as indicated in equation~(\ref{eq:mfp}). If this is  the case, the \AZ\ enhancement effect we describe increases the injection and acceleration efficiency of high \AZ\ particles compared to low \AZ\ ones. This dramatically decreases the abundance of accelerated electrons compared to heavier 
ions (e.g., Fig.~\ref{fig:PF_NL_gam10}).
The kinematics suggest that \rel\ shocks will not be able to place a sizable fraction of the shock kinetic energy flux into leptons if protons are present.

Of course, beyond kinematics, the magnetic turbulence produced by wave-particle interactions plays a critical role and recent results 
\citep[e.g.,][]{SironiSpit2011,Kumar2015} show that some fraction of the proton energy can be transferred to electrons in the shock precursor via magnetic turbulence. 
These PIC results are particularly important for astrophysical applications where the radiating electrons presumably contain a sizable fraction of the available energy budget. In  \rel\ shocks, heavy elements must transfer a sizable fraction of their energy to electrons for \Facc\ to be relevant for electrons.

We have modeled this energy transfer by including a parameter, $\EnTran$, that sets the fraction of ion energy transferred to electrons as the particles first cross the subshock. 
While the effect of $\EnTran$ is large 
(e.g., Fig.~\ref{fig:UM_gam10}), kinematics must still play a role:  light and heavy particles will be treated  differently in \rel\ flows. This is seen clearly in 
Figs.~\ref{fig:PF_NL_gam10} and \ref{fig:gam1p5_dNdp} where, for a given $\EnTran$, the $e/p$ ratio drops substantially between the UM case (where no \AZ\ effect occurs) and the NL case.
Figs.~\ref{fig:PF_NL_gam10} and 
\ref{fig:gam1p5_dNdp}, where $\EnTran=0.1$ for both the $\gamZ=10$ and $1.5$ shocks, also show that a larger fraction of ion energy must be transferred to electrons for high \Lor\ factor shocks to produce a significant $e/p$ ratio. 

We make a clear prediction that is directly testable  with PIC simulations. If \Facc\ is efficient enough so the shock structure is modified by the backpressure of accelerated particles, heavy element ions will show a clear enhancement over protons (i.e., Fig.~\ref{fig:dndp_AZ}). We know of no non-kinematic effects (e.g., cross-shock potentials, energy transfer via wave-particle interactions, or other electrostatic processes)  that can produce such an enhancement.

The combined processes of energy transfer from heavy particles to electrons, and the kinematics of shock smoothing, produce strong signatures on the radiation emitted by these particles. 
In Fig.~\ref{fig:phot_gam1p5} we show results for $\gamZ=10$ and $\gamZ=1.5$. 
Since there are a number of important parameters that influence the emission, such as shock \Lor\ factor, ambient density, magnetic field, and size of the emitting region, it is non-trivial to characterize the emission. Nevertheless some general properties stem mainly from the kinematics and should hold regardless of the plasma physics details.

Particle spectra should harden as the shock speed decreases from fully \rel\ to \nonrel\ speeds, mainly because the shock compression ratio increases and, for low enough $\gamZ$, $\Rtot > \rRH$ 
\citep[see figure~10 in][]{EWB2013}.
However, even though the compression ratio 
(defined as $\Rtot=u_0/u_2$) is lower for \ultrarel\ shocks, the downstream local plasma number density
$n_2 = \gamZ \betaZ n_0/(\gamT \betaT)$
can be large, enhancing the \pion\ emission more than IC 
and \synch.
The possibility of a significant change in the character of  
\NL\ effects  in the \transrel\ regime, as well as the fact that \transrel\ shocks have been observed \citep[e.g.,][]{Soderberg2010}, makes this an important area for future work.

The magnetic field is a critical parameter for \synch\ emission and, with the exception of Model \cCC, we have assumed $B_0=100$\,\muG\ for the background field. In our plane-parallel approximation, the background field remains constant throughout the shock.
Values of 100's of \muG\ can be expected for a shock moving through an ambient field of a few \muG\ when compression and amplification are considered.
Compression will increase the field by a factor $\sim \gamZ$ and NL amplification, as believed to occur in strong, \nonrel\ shocks in young SNRs \citep[see, for example,][ and references therein]{Bell2004,VBE2009,Bykov3inst2014}, may increase the strength further. 
For simplicity, we have not attempted to included compression or magnetic field amplification of the magnetic field here. 
Field compression is included in 
\citet{Warren2015dis}.

One important aspect of the changing afterglow emission as the shock slows from \ultrarel\ to \nonrel\ speeds is the position of the \synch\ peak. As seen in Fig.~\ref{fig:phot_gam1p5}, the peak shifts from $\sim$MeV to $\sim$keV as the shock slows from $\gamZ=10$ to $\gamZ=1.5$.
A detailed evolutionary model of GRB afterglows using \mc\ techniques for NL \Facc\ is presented in \citet{Warren2015dis}. In this afterglow model, the \mc\ simulation is combined with an analytic or numerical description of the jet-shock evolution. The shock accelerated particles and resultant radiation are calculated at various times as the shock moves through the jet and the total emission observed  at Earth is determined. 

We caution that, for simplicity, we have assumed Bohm diffusion 
here (i.e., equation~\ref{eq:mfp_two}; $\Lmfp = \rg \propto p$) whereas the actual scattering process in \rel\ plasmas is certain to be more complicated \citep[e.g.,][]{LP2010}. In particular, particles interacting with the small-scale turbulence generated by the Weibel instability are more likely to have $\Lmfp \propto p^2$ and additional instabilities may contribute longer scale turbulence with a different momentum dependence 
\citep[e.g.,][]{Casse2013,LPGP2014}. In fact, the momentum dependence and normalization of $\Lmfp$ can be expected to vary with momentum as well as position relative to the subshock. 
While a first-principles determination of $\Lmfp$ will undoubtedly require PIC  simulations, the \NL\ effects we describe here stem from basic considerations of momentum and energy conservation and should persist if first-order Fermi acceleration is efficient and $\Lmfp$ is an increasing function of $p$. 
Bearing in mind our simple scattering assumptions, we believe this work is the first to include electrons, protons, and heavier elements in a nonlinear relativistic shock acceleration model. We predict an \AZ\ enhancement effect for heavy ions in relativistic shock acceleration, and include the photon emission consistently with \NL\ particle acceleration. 
Nevertheless, modifying the scattering prescription is certain to produce  important quantitative differences, particularly for the photon emission which is strongly dependent on the maximum CR energy, and we are  generalizing the \mc\ technique to include more  realistic forms for $\Lmfp$ which will be presented in future work.

\acknowledgments The authors acknowledge useful comments by an anonymous referee and wish to thank Hirotaka Ito, John Kirk, Davide Lazzati, 
Martin Lemoine, Guy Pelletier, Steve Reynolds, and Lorenzo Sironi for helpful discussions. 
D.C.E. and D.C.W. acknowledge support from NASA
grant NNX11AE03G. A.M.B. was partially supported by the RAS Presidium Programm and the RAS OFN Programm n15. 
S-H.L acknowledges support from a JAXA International Top Young Fellowship. 
D.C.E. and S.-H.L. wish to thank the International Space Science Institute (ISSI) in Bern where part of this  work was done.

% bbbb  Note: must have files:  aa.bst  and  aa.cls
\bibliographystyle{aa} % A&A style
%\bibliography{c:/a_a_TOP/bibTeX/bib_DCE}
\bibliography{bib_DCE}

\clearpage

% Next sets line spacing from here to \end{spacing}
%\begin{spacing}{2.2}

% tttt
\begin{table}
\begin{center}
\caption{Model Parameters.}
\label{tab:param}
\renewcommand{\arraystretch}{1.5} % Changes line spacing in table
\begin{tabular}{crrrrrrrrrrrrrr}
\\
\tableline \tableline
\\
%
%%%
Model\tablenotemark{a}\tablenotetext{1}{Models \aaT, \bBB, and \cCC\ 
have $n_p=n_e = 1$\,\pcc\ with no helium. Models \dDD---\hHH\ have $n_p=1$\,\pcc, $\nHe=0.1$\,\pcc, and $n_e=1.2$\,\pcc. Models \iII\ and \jJJ\ have $n_p=1\,$\pcc\ and $\nHe=0.1 n_p$ without electrons. The far upstream temperature is $10^6$\,K in all cases and all models have $\etamfp=1$.}
&Type\tablenotemark{b}\tablenotetext{2}{In the \NL\ (NL) models the shock structure is determined \SCly. The \UM\ (UM) models have a discontinuous shock structure with no shock smoothing.}
&$\gamZ$ 
&$\EnTran$ 
&$B_0$ 
&$\LfebUpS$
&$\LfebDS$
&$\rRH$
&$\Rtot$
&$N_g$
&$\EffGeVp$\tablenotemark{c}\tablenotetext{3}{For \SC\ NL models, this is the fraction of total energy placed in all particles with energies 
above 100\,MeV as measured in the shock frame.} 
&$\EffGeVhe$
&$\EffGeVe$
\\
\tableline
&&&
&[\muG]
&$\rgz$
&$\rgz$\\
\tableline
\aaT 
&UM
&10 
&0 
&100 
&$10^4$ 
&\dots 
&3.02 
&3.02 
&2000
&\dots
&\dots
&\dots
\\
\bBB 
&UM 
&10 
&0.15 
&100 
&$10^4$ 
&\dots 
&3.02
&3.02 
&2000
&\dots
&\dots
&\dots
\\
\cCC
&UM 
&10 
&0.15 
&3 
&$10^3$ 
&$10^3$
&3.02
&3.02 
&2000
&\dots
&\dots
&\dots
\\
\dDD
&UM 
&10 
&0.1 
&100
&$300$ 
&$10^3$
&3.02
&3.02 
&2000
&\dots
&\dots
&\dots
\\
\eEE
&NL 
&10 
&0.1 
&100
&$300$ 
&$10^3$
&3.02
&3.02 
&2000
&0.60
&0.30
&0.10
\\
\fFF
&NL 
&10 
&0.4 
&100
&$300$ 
&$10^3$
&3.02
&3.02 
&2000
&0.36
&0.20
&0.44
\\
\gGG
&UM 
&1.5
&0.1 
&100
&$300$ 
&$10^3$
&3.53
&3.53 
&200
&\dots
&\dots
&\dots
\\
\hHH
&NL 
&1.5
&0.1 
&100
&$300$ 
&$10^3$
&3.53
&$3.9\pm 0.4$ 
&200
&0.52
&0.27
&0.08
\\
\iII
&UM 
&30
&\dots 
&100
&$300$ 
&$10^4$
&3.00
&3.00 
&$10^4$
&\dots
&\dots
&\dots
\\
\jJJ
&NL
&30
&\dots 
&100
&$300$ 
&$10^4$
&3.00
&3.00 
&$10^4$
&0.66
&0.34
&\dots
\\
\tableline
\end{tabular}
\end{center}
\end{table}
%above tttt

%\end{spacing}

\end{document}